# Comparing and Scaling fMRI Features for Brain-Behavior Prediction

Mikkel Schöttner[1*], Thomas A.W. Bolton[1], Jagruti Patel[1], Patric Hagmann[1]

[1] Department of Radiology, Lausanne University Hospital and University of Lausanne (CHUV-UNIL), Lausanne, Switzerland
* Email: mikkel.schottner@unil.ch

## Abstract

Predicting behavioral variables from neuroimaging modalities such as magnetic resonance imaging (MRI) has the potential to allow the development of neuroimaging biomarkers of mental and neurological disorders. A crucial processing step to this aim is the extraction of suitable features. These can differ in how well they predict the target of interest, and how this prediction scales with sample size and scan time. Here, we compare nine feature subtypes extracted from resting-state functional MRI recordings for behavior prediction, ranging from regional measures of functional activity to functional connectivity (FC) and metrics derived with graph signal processing (GSP), a principled approach for the extraction of structure-informed functional features.

We study 979 subjects from the Human Connectome Project Young Adult dataset, predicting summary scores for mental health, cognition, processing speed, and substance use, as well as age and sex. The scaling properties of the features are investigated for different combinations of sample size and scan time.

FC comes out as the best feature for predicting cognition, age, and sex. Graph power spectral density is the second best for predicting cognition and age, while for sex, variability-based features show potential as well. When predicting sex, the low-pass graph filtered coupled FC slightly outperforms the simple FC variant. None of the other targets were predicted significantly. The scaling results point to higher performance reserves for the better-performing features. They also indicate that it is important to balance sample size and scan time when acquiring data for prediction studies.

The results confirm FC as a robust feature for behavior prediction, but also show the potential of GSP and variability-based measures. We discuss the implications for future prediction studies in terms of strategies for acquisition and sample composition.

# 1 Introduction

Behavior prediction with the goal to develop neuroimaging biomarkers of mental illness has become an increasingly important area in neuroscience, as it might allow for more individualized treatment and opens the door to early detection and treatment selection based on the underlying biology (Bzdok & Meyer-Lindenberg, 2018; Dubois & Adolphs, 2016; Finn & Todd Constable, 2016; Gabrieli et al., 2015; Lui et al., 2016). Following the idea of precision medicine, neuroimaging biomarkers offer to be a tool for efficient and robust early detection of disorders. Prior studies that investigated how to predict behavioral variables from neuroimaging data have largely focused on functional connectivity (FC) and its use as a feature to predict cognitive abilities (He et al., 2020; Ooi et al., 2022; Vieira et al., 2022). Here, we compare the ability of various measures of brain activity to predict several behavioral dimensions (Schöttner et al., 2023), as well as age and sex. We combine measures with established effects in group comparisons with those derived from novel structure-informed features, and FC as the benchmark in one comparative study.

While FC has shown the best results so far, other fMRI features could be considered for behavior prediction, by combining other modalities with fMRI and using more complex analysis methods. FC, as a connection-wise measure, does not allow for direct inference of the brain regions involved, because of the machine learning model that needs to be used to handle its high dimensionality. Regional measures of brain activity might thus be a more efficient representation that also allows for these inferences. Different types of features might also be better at predicting other targets than cognition.

Several regional measures of fMRI activity, which we considered in this study, have shown potential in past works. Simply taking the mean and standard deviation has been successfully employed to classify treatment outcome in social anxiety disorder (Månsson et al., 2022) and autism spectrum disorder (ASD, Brahim & Farrugia, 2020), the latter of which was improved by using GSP (see below). Moment-to-moment changes in BOLD variability (Garrett et al., 2013), as measured through the mean square successive difference (MSSD), have been related to cognition and age (Boylan et al., 2021; Nomi et al., 2017), and were found to be different between schizophrenia patients and healthy controls in specific frequency bands (Zhang et al., 2021). Finally, the fractional amplitude of low-frequency fluctuations (fALFF) of the brain signal (Q.-H. Zou et al., 2008) was shown to be related to cognitive impairment in the Alzheimer's disease spectrum, and different between major depressive disorder patients and controls (Lai & Wu, 2015; Yang et al., 2019); it was also used to classify schizophrenia patients from healthy controls (Savio & Graña, 2015).

While these measures offer simple ways to compute novel features, another promising approach makes use of GSP, a set of tools that leverage the network structure of the brain and process signals that can be represented on top of this structure (for a review of its applications in neuroimaging, see Lioi et al., 2021). In addition to improving prediction of ASD from mean and standard deviation using the graph Fourier transform (Brahim & Farrugia, 2020), multiple works have shown the potential of GSP in task decoding (Ménoret et al., 2017; Pilavci & Farrugia, 2019; Xu et al., 2019). Furthermore, several new metrics have been introduced in the neuroimaging context. The structural decoupling index (SDI, Preti & Van De Ville, 2019) describes how much the fMRI signal is constrained by the underlying scaffolding of structural connectivity (SC). It is based on the relative strength of the signal that was filtered using a graph high-pass or low-pass filter, and was successfully employed for task decoding and fingerprinting (Griffa et al., 2022). These filtered signals can also be used to compute coupled and decoupled FC matrices (Griffa et al., 2022), respectively from the low- and high-pass filtered signals. Finally, computing the power spectral density of each graph frequency was also considered as a feature in this study, which has been shown to change under the acute influence of psychedelic drug use (Atasoy et al., 2017).

While comparing these features in their ability to predict different behavioral variables is interesting in itself, the scaling properties of features are also an important consideration in the development of neuroimaging biomarkers. Investigating how prediction performance changes in relation to sample size and scan time can give an indication of which features have performance reserves that can be exploited in larger datasets, and which features are already at their maximum performance. Moreover, choosing to scan more subjects or each subject for longer is also an economical consideration, posing a tradeoff which has been explored by Ooi et al. (2024) for FC. In order to make economically viable decisions when it comes to planning the acquisition schemes of studies, especially large ones, it is important to know whether it will pay off more to add more subjects, or to scan subjects for longer.

The goal of this study is thus to investigate the relative effectiveness of different MRI measures in predicting behavior, and how prediction performance of these features scales with subjects and scanning time. Prediction of behavioral variables is the first step to develop neuroimaging biomarkers of mental illness. However, due to the multitude of methodological choices one can make in brain-behavior prediction (Dhamala et al., 2023), it is important to evaluate how each of these choices influence prediction accuracy. One such choice is how to extract features from the neuroimaging data. For this reason, this study offers a methodological comparison of different fMRI features for prediction of behavior.

# 2 Methods

## 2.1 Dataset & Preprocessing

We used 979 subjects from the Human Connectome Project (HCP) Young Adult dataset (Van Essen et al., 2012) with complete structural and diffusion MRI and resting-state fMRI. Participants gave informed consent, and all recruitment and acquisition methods were approved by the Washington University Institutional Review Board (IRB), following all relevant guidelines and regulations. Participants were between 22 and 37 years old ($28.68 \pm 3.71$); 525 were female, 454 were male.

The structural T1 images were processed using Connectomemapper3 v.3.0.0-rc4 (Tourbier et al., 2021, 2022), which combines different processing tools into a pipeline. As part of this pipeline, the images were parcellated using the Lausanne 2018 atlas at scale 3, which comprises $R = 274$ regions, including subcortical structures, cerebellum, and brainstem (Cammoun et al., 2012; Desikan et al., 2006; Iglesias, Augustinack, et al., 2015; Iglesias, Van Leemput, et al., 2015; Najdenovska et al., 2018).

The diffusion images were also processed using Connectomemapper3, using MRtrix (Tournier et al., 2012) with constrained spherical deconvolution and deterministic tractography with white matter seeding and 10 million streamlines. T1-weighted and diffusion images were then combined to create structural connectivity (SC) matrices, with normalized fiber density as the edge weight and self-connections set to zero.

We used the minimally processed functional images, which were already treated with distortion correction, realignment, coregistration to structural image, bias field correction, normalization, and masked with a brain mask (Glasser et al., 2013). After discarding the first 6 images to get rid of scanner drift, we additionally performed confound regression, using six motion parameters and their first-order derivatives, detrending, and high-pass filtering, at a cutoff of 0.01 Hz. The fMRI time series were parcellated using the Lausanne 2018 atlas by averaging the signal over all voxels in each parcel.

As behavioral prediction targets, we used four summary scores that were found using exploratory factor analysis, using the same procedure as described in Schöttner et al. (2023). In order to avoid data leakage, the factor scores were derived on a subset of 145 subjects, leaving 834 subjects for the main analysis. The factors found were the same as in our earlier work, namely mental health, cognition, processing speed, and substance use. Additionally, the age in years and the sex of participants were also used as prediction targets, as these provide ground truth measurements.

## 2.2 fMRI Features

### 2.2.1 Functional Connectivity

As the baseline feature we considered functional connectivity, shown to perform well in previous comparisons of features in brain-behavior prediction studies (Ooi et al., 2022). For this, the Pearson correlation coefficient between the time series of each pair of regions was calculated. The upper triangle of the FC matrix was extracted and vectorized, resulting in a vector of dimension $\frac{R \times (R-1)}{2} = 37401$.

### 2.2.2 Region-Wise fMRI Features

Next, we considered some features that were calculated only using the fMRI time series, which we grouped as region-wise features. These were measures that have previously been shown to relate to behavior, mostly in classification and group-comparison studies. The simplest of these were the mean and standard deviation (SD) of each time series. The mean squared successive difference (MSSD), also called BOLD variability in the literature (Garrett et al., 2013), quantifies the moment-to-moment change in signal. It is calculated as

$$MSSD = \frac{1}{N-1} \sum_{i=1}^{N-1} (x_{i+1} - x_i)^2$$

where $N$ is the number of time points in the fMRI time series and $x_i$ is the value of the $i$th data point. Finally, the fractional amplitude of low-frequency fluctuations (fALFF) is defined as the ratio of signal power in the low-frequency range (0.01-0.08 Hz) to the signal power in the whole frequency range (Q.-H. Zou et al., 2008). As all these measures were regional measures, they resulted in feature vectors of dimension $R = 274$.

### 2.2.3 Graph Signal Processing Features

Some measures were calculated with the help of graph signal processing (GSP), combining structural coupling information with fMRI signals. For this, we first created a common structural connectivity matrix. We created a binary version that thresholded the connection weights, preserving the edge distribution, which can otherwise get biased towards short connections (details to this approach in Betzel et al., 2019). A simple average of connection strengths over all considered subjects was also computed. The Hadamard product gave the common structural connectivity matrix, which both preserved the edge distribution and edge weights. In order to avoid data leakage, this matrix was recomputed for each training set (see below). Then, we performed an eigenvalue decomposition of the normalized Laplacian of this matrix to get the structural connectome harmonics. The graph Fourier transform was used to convert the regional fMRI signal to the graph domain, meaning that it was now represented as a weighted sum of connectome harmonics. The PSD was obtained by calculating the second order norm of the squared signal in the graph domain over time. Analogous to Griffa et al. (2022), we then created high- and low-pass filtered versions of the fMRI signal, by setting the graph coefficients below and above a threshold to zero, respectively. Instead of defining the threshold based on the relative energy, we opted to instead use $\frac{R}{2}$ as our cutoff frequency, as previous analyses showed that this worked better for behavior prediction than splitting by energy (Schöttner et al., 2024). By computing the pair-wise correlations for the low-pass and high-pass filtered signals, we arrived at coupled and decoupled FC matrices, respectively. As with the regular FC matrix, these were vectorized, and only the upper triangle was used as features. Finally, the structural decoupling index (SDI) was calculated for each region, which is defined as the ratio between the norms of the high- and low-pass filtered signals per region (Preti & Van De Ville, 2019). The dimensions of the GSP features were thus $R = 274$ for PSD and SDI, and $\frac{R \times (R-1)}{2} = 37401$ for coupled and decoupled FC.

### 2.3 Behavior Prediction Setup

In total, 9 different features were compared in this study: FC, mean and SD of the fMRI signals, BOLD variability, fALFF, PSD, SDI, as well as coupled and decoupled FC. The targets were the 4 factor scores: mental health, cognition, processing speed, substance use, as well as age and sex. The prediction pipeline was the following: within the training set, we split the data into 10 random train/test splits, combined with nested cross-validation. The size of the test set was

15 % of the whole training set. Because the families needed to be kept in the same split, the train and test sets differed in size, ranging between 693 and 720, and 114 and 141, respectively. There were 3 inner folds to optimize the hyperparameters. Features were scaled using a standard scaler.

For the continuous targets, we used two different models: elastic-net and kernel ridge regression (KRR). Elastic-net is a linear regression method that uses a combination of L1 and L2 regularization, making the solution sparse and the weights small (H. Zou & Hastie, 2005). KRR uses the kernel trick in order to handle high dimensional features, and has been successfully used for this task before (Ooi et al., 2022). Both models were fit for all targets. For the discrete target sex, we used an elastic-net classifier and a support vector machine (SVM), which we deemed the closest equivalents to the continuous models. All continuous models were evaluated using the coefficient of determination ($R^2$), the discrete models using accuracy.

In order to test whether the features predict the target above chance, we employed a permutation procedure to generate a null distribution to compare the predictions against. For this, the target scores were reshuffled 100 times for each of the 10 train test splits and feature target combination. Following that, the prediction models were fit using the same procedure as described above. This resulted in a null distribution of 1000 $R^2$ values for the reshuffled targets. To test each feature for significance, the mean of the $R^2$ distribution predicting the real targets was compared against the 95th percentile of the reshuffled $R^2$ distribution.

## 2.4 Scaling

To investigate scaling effects, we repeated the same prediction pipelines with different combinations of fractions of the training set and scanning sessions. The fractions of the training set were ranging between 0.2 and 1, in increments of 0.2. As mentioned above, because members of the same family had to be contained in the same split, the splits did not always have the same size, and are thus just expressed as fractions of the training set instead of discrete numbers. The size of the test set was kept constant for the different fractions of the training set. The fractions of the scanning sessions started at 0.25 (3.6 min), doubling in length until all four sessions were used (57.6 min). To control for effects unique to the start of the session, the session fractions always started with the beginning of session one, and were extended from there.

# 3 Results

## 3.1 FC and Coupled FC Are the Best Features

Tables 1 and 2 show the prediction results for the continuous targets mental health, cognition, processing speed, substance use, and age, when using elastic-net regression or KRR, respectively. For both models, only cognition and age could be predicted higher than chance level, i.e., only for those targets was the distribution of $R^2$ values significantly different from the null-distribution as indicated by the permutation test. For those targets, all features had significant predictions except when using the mean. Table 3 shows the prediction results for sex, for both models (elastic-net regression and SVM). For either model, all features except mean yielded significant predictions. Figures 1-3 show how well cognition, age, and sex were predicted by each of the significant features using all available data. The figures additionally show the significant features' respective scaling curves over the number of training subjects and scanning sessions.

When predicting cognition, the FC-based features did best, with FC taking the top spot when using KRR, and coupled FC when using elastic-net regression, followed by decoupled FC for either model. The features with dimensionality R showed lower performance, here PSD (for KRR) and standard deviation (for elastic-net) came out ahead, followed by SDI, fALFF, and BOLD variability. Taken together, using all data, both models performed comparably, with only slight differences depending on the feature. The scaling curves of the best features look like they are saturating. For KRR, it seems like there is still an upwards trend, indicating that there are performance reserves for larger sample sizes and scanning times, at least for FC. Worse features saturated earlier, which could indicate that performance reserves only exist for the better-performing features.

**Table 1**
*Prediction of continuous targets using elastic-net regression.*

| Target | Feature | Mean | Standard Deviation | Median | Q1 | Q3 | p-value |
|---|---|---|---|---|---|---|---|
| Mental Health | Mean | -0.032 | 0.036 | -0.017 | -0.066 | -0.002 | 0.924 |
| | Standard Deviation | -0.031 | 0.036 | -0.013 | -0.065 | -0.003 | 0.926 |
| | BOLD Variability | -0.031 | 0.036 | -0.013 | -0.065 | -0.003 | 0.917 |
| | fALFF | -0.033 | 0.035 | -0.017 | -0.065 | -0.003 | 0.92 |
| | FC | -0.031 | 0.036 | -0.013 | -0.065 | -0.003 | 0.919 |
| | Structural Decoupling Index | -0.036 | 0.04 | -0.013 | -0.077 | -0.003 | 0.94 |
| | Power Spectral Density | -0.031 | 0.036 | -0.013 | -0.065 | -0.003 | 0.911 |
| | Coupled FC | -0.034 | 0.034 | -0.017 | -0.065 | -0.007 | 0.94 |
| | Decoupled FC | -0.031 | 0.036 | -0.013 | -0.065 | -0.003 | 0.923 |
| Cognition | Mean | -0.023 | 0.022 | -0.02 | -0.03 | -0.008 | 0.849 |
| | Standard Deviation | 0.049 | 0.064 | 0.058 | -0.008 | 0.108 | <0.001 |
| | BOLD Variability | 0.01 | 0.086 | 0.046 | -0.048 | 0.061 | 0.007 |
| | fALFF | 0.027 | 0.053 | 0.023 | -0.008 | 0.054 | <0.001 |
| | FC | 0.122 | 0.104 | 0.139 | 0.102 | 0.194 | <0.001 |
| | Structural Decoupling Index | 0.035 | 0.038 | 0.038 | 0.017 | 0.058 | 0.001 |
| | Power Spectral Density | 0.038 | 0.086 | 0.055 | -0.048 | 0.118 | <0.001 |
| | Coupled FC | 0.138 | 0.068 | 0.147 | 0.116 | 0.184 | <0.001 |
| | Decoupled FC | 0.09 | 0.073 | 0.084 | 0.034 | 0.147 | <0.001 |
| Substance Use | Mean | -0.03 | 0.034 | -0.018 | -0.035 | -0.008 | 0.908 |
| | Standard Deviation | -0.024 | 0.036 | -0.014 | -0.024 | -0.003 | 0.861 |
| | BOLD Variability | -0.017 | 0.035 | 0.003 | -0.019 | 0.003 | 0.776 |
| | fALFF | -0.028 | 0.034 | -0.016 | -0.038 | -0.003 | 0.888 |
| | FC | -0.026 | 0.037 | -0.012 | -0.024 | -0.002 | 0.867 |
| | Structural Decoupling Index | -0.016 | 0.03 | -0.003 | -0.033 | 0.007 | 0.772 |
| | Power Spectral Density | -0.028 | 0.04 | -0.015 | -0.022 | -0.003 | 0.866 |
| | Coupled FC | -0.036 | 0.045 | -0.015 | -0.044 | -0.006 | 0.918 |
| | Decoupled FC | -0.016 | 0.036 | -0.007 | -0.02 | 0.011 | 0.794 |
| Processing Speed | Mean | -0.011 | 0.018 | -0.006 | -0.01 | -0.001 | 0.661 |
| | Standard Deviation | -0.014 | 0.018 | -0.008 | -0.017 | -0.001 | 0.727 |
| | BOLD Variability | -0.009 | 0.02 | -0.003 | -0.012 | 0.003 | 0.596 |
| | fALFF | -0.003 | 0.024 | 0.005 | -0.017 | 0.014 | 0.372 |
| | FC | -0.013 | 0.017 | -0.007 | -0.015 | -0.003 | 0.759 |
| | Structural Decoupling Index | -0.011 | 0.018 | -0.004 | -0.011 | -0.002 | 0.684 |
| | Power Spectral Density | -0.013 | 0.018 | -0.007 | -0.016 | -0.003 | 0.711 |
| | Coupled FC | -0.009 | 0.018 | -0.003 | -0.008 | -0.002 | 0.638 |
| | Decoupled FC | -0.013 | 0.018 | -0.007 | -0.017 | -0.003 | 0.718 |
| Age | Mean | -0.015 | 0.014 | -0.014 | -0.022 | -0.003 | 0.768 |
| | Standard Deviation | 0.115 | 0.055 | 0.118 | 0.069 | 0.153 | <0.001 |
| | BOLD Variability | 0.085 | 0.043 | 0.066 | 0.061 | 0.098 | <0.001 |
| | fALFF | 0.112 | 0.041 | 0.123 | 0.099 | 0.137 | <0.001 |
| | FC | 0.22 | 0.052 | 0.213 | 0.201 | 0.239 | <0.001 |
| | Structural Decoupling Index | 0.145 | 0.042 | 0.144 | 0.114 | 0.174 | <0.001 |
| | Power Spectral Density | 0.181 | 0.034 | 0.171 | 0.155 | 0.199 | <0.001 |
| | Coupled FC | 0.19 | 0.052 | 0.19 | 0.147 | 0.235 | <0.001 |
| | Decoupled FC | 0.154 | 0.041 | 0.146 | 0.134 | 0.171 | <0.001 |

*Note.* Mean, standard deviation, median, first quartile, third quartile, and p-value of the $R^2$ performance over 10 splits for each target and feature combination.

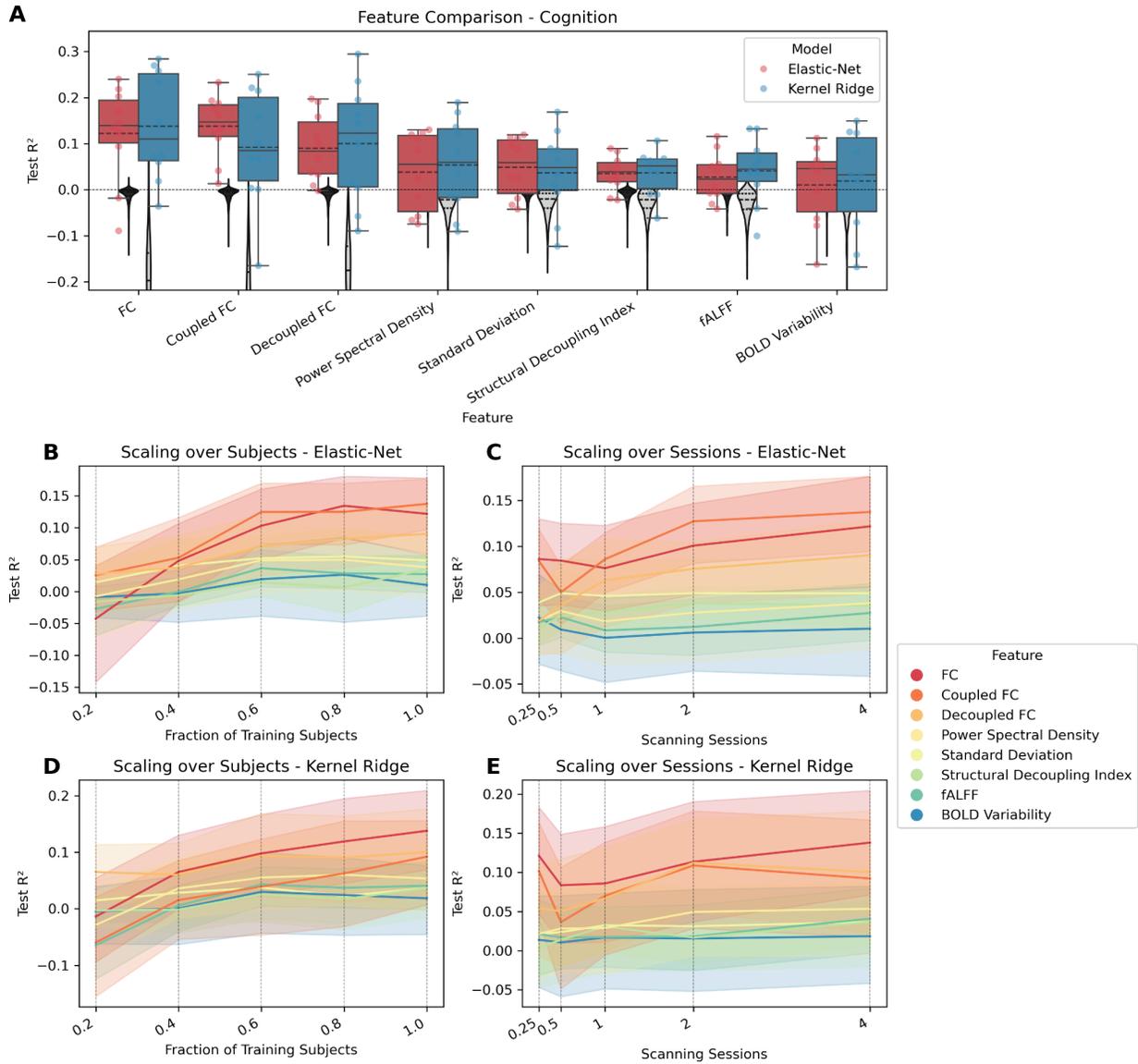

*Figure 1: Comparing features for predicting cognition. FC variants have dimension $\frac{R \times (R-1)}{2} = 37401$, while all other features have dimension $R = 274$. (A) Coefficient of determination (R²) for each feature that differs significantly from zero. Features are ordered by mean. Each point represents the performance of one split. The distribution is shown as a boxplot, where the solid line represents the median and the dashed line the mean. The zero line is plotted as a horizontal dotted line. The null distribution is shown as a gray violin plot, the dashed line representing mean and quartiles. (B) Lineplot showing test performance (R²) over fractions of training subjects. Each vertical dotted line shows a value on the x-axis where the data are plotted. (C) Lineplot showing test performance (R²) over scanning sessions. Each vertical dotted line shows a value on the x-axis where the data are plotted. The surfaces represent the 95 % confidence interval.*

For age, again, FC-based measures were among the best, regular FC coming out ahead. PSD also did relatively well, slightly better than decoupled FC, but lower than the other two FC variants. SDI and variability based measures showed relatively worse performance. Additionally, the difference between models is more striking when predicting age than for cognition, with KRR yielding higher R² values than elastic-net. In terms of the scaling curves, there was less leveling off for the well-performing features than it was the case for cognition, indicating that with more subjects or longer scanning time, predictions could be improved further. A factor that might play a role here is the small age range across HCP subjects, meaning that instead of adding more subjects or scanning subjects for a longer time, a more diverse sample regarding age could also improve performance.

**Table 2**
*Prediction of continuous targets using KRR.*

| Target | Feature | Mean | Standard Deviation | Median | Q1 | Q3 | p-value |
|---|---|---|---|---|---|---|---|
| Mental Health | Mean | -0.019 | 0.024 | -0.021 | -0.037 | 0.001 | 0.769 |
| | Standard Deviation | -0.029 | 0.024 | -0.021 | -0.053 | -0.008 | 0.886 |
| | BOLD Variability | -0.027 | 0.025 | -0.02 | -0.052 | -0.005 | 0.855 |
| | fALFF | -0.031 | 0.033 | -0.031 | -0.052 | -0.005 | 0.891 |
| | FC | -0.275 | 0.087 | -0.271 | -0.352 | -0.198 | 0.838 |
| | Structural Decoupling Index | -0.032 | 0.033 | -0.031 | -0.059 | -0.001 | 0.899 |
| | Power Spectral Density | -0.03 | 0.026 | -0.021 | -0.049 | -0.009 | 0.873 |
| | Coupled FC | -0.349 | 0.085 | -0.34 | -0.431 | -0.291 | 0.877 |
| | Decoupled FC | -0.177 | 0.061 | -0.16 | -0.225 | -0.125 | 0.565 |
| Cognition | Mean | -0.019 | 0.045 | -0.002 | -0.014 | 0.003 | 0.472 |
| | Standard Deviation | 0.037 | 0.091 | 0.048 | -0.001 | 0.089 | <0.001 |
| | BOLD Variability | 0.019 | 0.112 | 0.032 | -0.048 | 0.113 | 0.004 |
| | fALFF | 0.041 | 0.072 | 0.044 | 0.018 | 0.079 | <0.001 |
| | FC | 0.138 | 0.116 | 0.11 | 0.063 | 0.252 | <0.001 |
| | Structural Decoupling Index | 0.036 | 0.05 | 0.051 | 0.003 | 0.066 | <0.001 |
| | Power Spectral Density | 0.054 | 0.101 | 0.059 | -0.017 | 0.132 | <0.001 |
| | Coupled FC | 0.092 | 0.127 | 0.085 | 0.019 | 0.201 | <0.001 |
| | Decoupled FC | 0.1 | 0.128 | 0.123 | 0.006 | 0.187 | <0.001 |
| Substance Use | Mean | -0.035 | 0.034 | -0.03 | -0.042 | -0.009 | 0.851 |
| | Standard Deviation | -0.021 | 0.047 | -0.006 | -0.031 | 0.002 | 0.718 |
| | BOLD Variability | -0.019 | 0.048 | -0.007 | -0.023 | 0.016 | 0.684 |
| | fALFF | -0.035 | 0.043 | -0.017 | -0.071 | -0.006 | 0.847 |
| | FC | -0.192 | 0.206 | -0.157 | -0.271 | -0.047 | 0.506 |
| | Structural Decoupling Index | -0.014 | 0.056 | 0.01 | -0.036 | 0.025 | 0.59 |
| | Power Spectral Density | -0.018 | 0.047 | -0.01 | -0.026 | 0.012 | 0.661 |
| | Coupled FC | -0.267 | 0.27 | -0.183 | -0.348 | -0.099 | 0.616 |
| | Decoupled FC | -0.134 | 0.195 | -0.026 | -0.266 | 0.005 | 0.374 |
| Processing Speed | Mean | -0.018 | 0.017 | -0.013 | -0.021 | -0.007 | 0.749 |
| | Standard Deviation | -0.017 | 0.019 | -0.021 | -0.026 | -0.003 | 0.683 |
| | BOLD Variability | -0.007 | 0.018 | -0.011 | -0.017 | 0.003 | 0.447 |
| | fALFF | -0.012 | 0.024 | -0.005 | -0.016 | 0.005 | 0.62 |
| | FC | -0.108 | 0.079 | -0.11 | -0.16 | -0.047 | 0.175 |
| | Structural Decoupling Index | -0.01 | 0.018 | -0.005 | -0.017 | 0.002 | 0.529 |
| | Power Spectral Density | -0.019 | 0.017 | -0.018 | -0.024 | -0.007 | 0.734 |
| | Coupled FC | -0.154 | 0.08 | -0.15 | -0.186 | -0.115 | 0.239 |
| | Decoupled FC | -0.132 | 0.131 | -0.107 | -0.171 | -0.063 | 0.318 |
| Age | Mean | -0.011 | 0.016 | -0.006 | -0.016 | -0.001 | 0.57 |
| | Standard Deviation | 0.145 | 0.057 | 0.153 | 0.096 | 0.167 | <0.001 |
| | BOLD Variability | 0.101 | 0.043 | 0.101 | 0.07 | 0.128 | <0.001 |
| | fALFF | 0.131 | 0.055 | 0.12 | 0.105 | 0.171 | <0.001 |
| | FC | 0.275 | 0.076 | 0.26 | 0.226 | 0.296 | <0.001 |
| | Structural Decoupling Index | 0.165 | 0.046 | 0.169 | 0.136 | 0.182 | <0.001 |
| | Power Spectral Density | 0.189 | 0.054 | 0.191 | 0.162 | 0.204 | <0.001 |
| | Coupled FC | 0.263 | 0.106 | 0.244 | 0.184 | 0.329 | <0.001 |
| | Decoupled FC | 0.173 | 0.057 | 0.17 | 0.135 | 0.188 | <0.001 |

*Note.* Mean, standard deviation, median, first quartile, third quartile, and p-value of the $R^2$ performance over 10 splits for each target and feature combination.

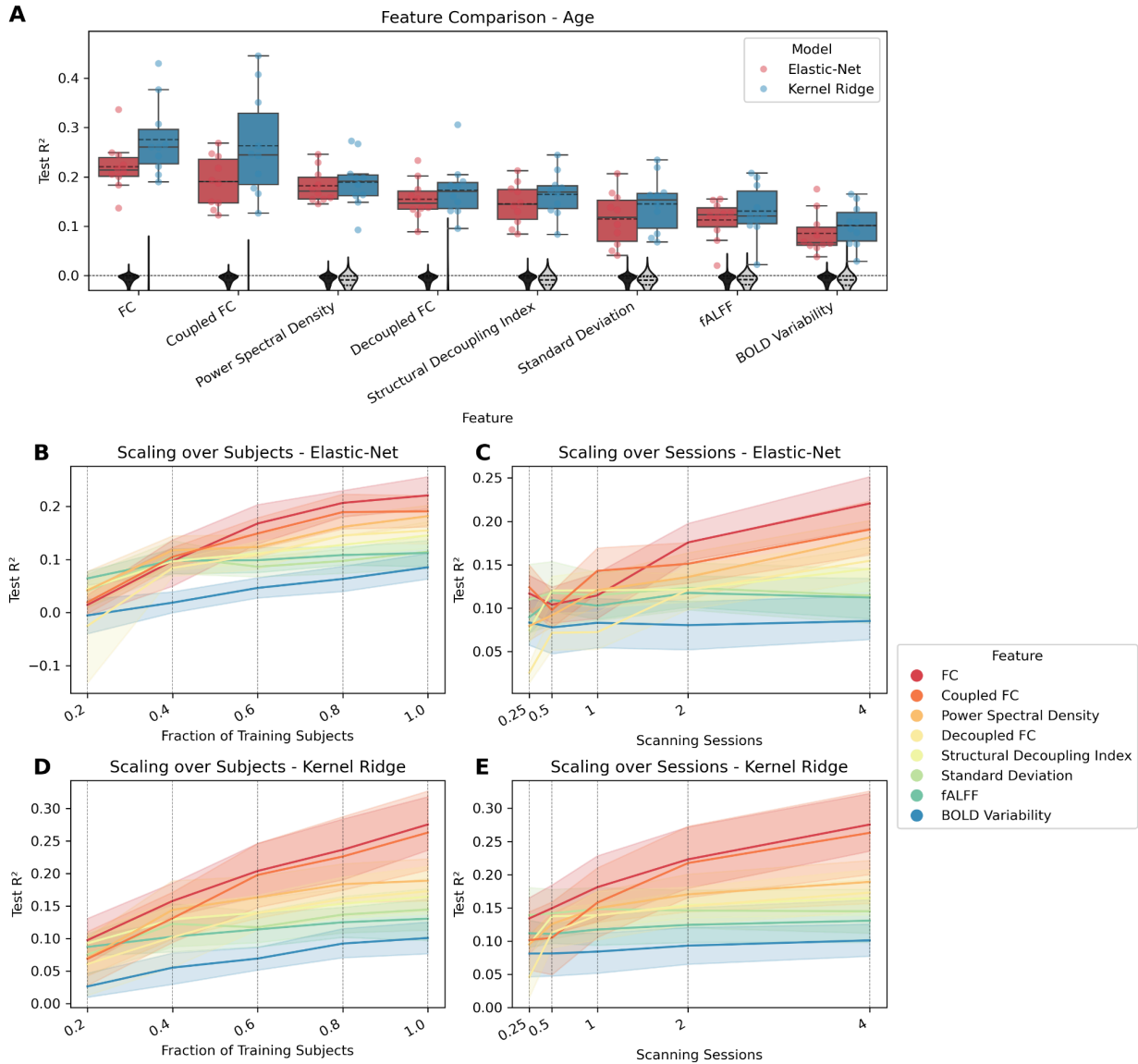

Figure 2: Comparing features for predicting age. FC variants have dimension $\frac{R \times (R-1)}{2} = 37401$, while all other features have dimension $R = 274$. (A) Coefficient of determination ($R^2$) for each feature that differs significantly from zero. Features are ordered by mean. Each point represents the performance of one split. The distribution is shown as a boxplot, where the solid line represents the median and the dashed line the mean. The zero line is plotted as a horizontal dotted line. The null distribution is shown as a gray violin plot, the dashed line representing mean and quartiles. (B) Lineplot showing test performance ($R^2$) over fractions of training subjects. Each vertical dotted line shows a value on the x-axis where the data are plotted. (C) Lineplot showing test performance ($R^2$) over scanning sessions. Each vertical dotted line shows a value on the x-axis where the data are plotted. The surfaces represent the 95 % confidence interval.

For sex, top features yet again included FC-based ones, but also measures of regional variability. As was also the case for when predicting cognition using elastic-net regression, coupled FC came out slightly ahead of using FC, in this case for both models. Standard deviation outperformed decoupled FC, which was followed by BOLD variability, PSD, SDI, and fALFF. For most features, using SVM as the model yielded higher prediction performances than the elastic-net classifier. The scaling curves leveled off similar to how they do for cognition, indicating that there are probably less performance reserves than for age. This is more so the case for SVM than for the elastic-net classifier, the former of which already had higher accuracy values.

**Table 3**
*Predicting sex.*

| Model | Feature | Mean | Standard Deviation | Median | Q1 | Q3 | p-value |
|---|---|---|---|---|---|---|---|
| Elastic-Net Classifier | Mean | 0.526 | 0.065 | 0.525 | 0.485 | 0.562 | 0.241 |
| | Standard Deviation | 0.817 | 0.048 | 0.82 | 0.793 | 0.846 | <0.001 |
| | BOLD Variability | 0.81 | 0.042 | 0.815 | 0.777 | 0.841 | <0.001 |
| | fALFF | 0.748 | 0.059 | 0.754 | 0.711 | 0.79 | <0.001 |
| | FC | 0.801 | 0.063 | 0.808 | 0.767 | 0.846 | <0.001 |
| | Structural Decoupling Index | 0.736 | 0.066 | 0.742 | 0.698 | 0.781 | <0.001 |
| | Power Spectral Density | 0.774 | 0.049 | 0.777 | 0.737 | 0.808 | <0.001 |
| | Coupled FC | 0.803 | 0.067 | 0.815 | 0.76 | 0.851 | <0.001 |
| | Decoupled FC | 0.782 | 0.056 | 0.785 | 0.745 | 0.823 | <0.001 |
| SVM | Mean | 0.51 | 0.039 | 0.508 | 0.484 | 0.537 | 0.312 |
| | Standard Deviation | 0.834 | 0.046 | 0.838 | 0.807 | 0.868 | <0.001 |
| | BOLD Variability | 0.824 | 0.038 | 0.824 | 0.802 | 0.854 | <0.001 |
| | fALFF | 0.768 | 0.052 | 0.776 | 0.738 | 0.802 | <0.001 |
| | FC | 0.83 | 0.057 | 0.837 | 0.794 | 0.87 | <0.001 |
| | Structural Decoupling Index | 0.76 | 0.06 | 0.769 | 0.716 | 0.805 | <0.001 |
| | Power Spectral Density | 0.785 | 0.046 | 0.786 | 0.754 | 0.815 | <0.001 |
| | Coupled FC | 0.826 | 0.061 | 0.831 | 0.784 | 0.87 | <0.001 |
| | Decoupled FC | 0.809 | 0.054 | 0.809 | 0.769 | 0.853 | <0.001 |

*Note.* Mean, standard deviation, median, first quartile, third quartile, and p-value of the $R^2$ performance over 10 splits for each model and feature combination.

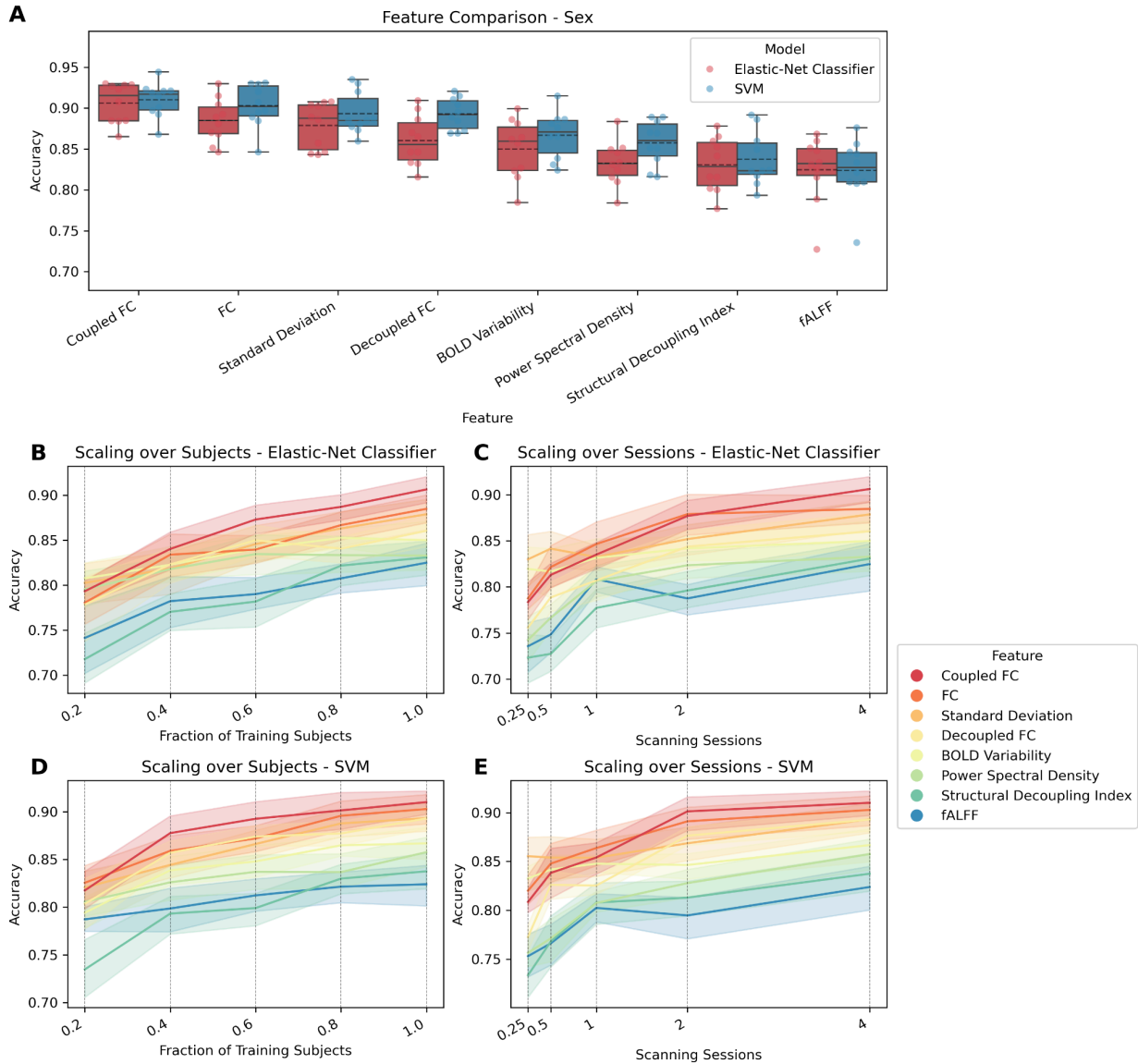

*Figure 3: Comparing features for predicting sex. FC variants have dimension $\frac{R \times (R-1)}{2} = 37401$, while all other features have dimension $R = 274$. (A) Accuracy for each feature that differs significantly from zero. Features are ordered by mean. Each point represents the performance of one split. The distribution is shown as a boxplot, where the solid line represents the median and the dashed line the mean. (B) Lineplot showing test performance (accuracy) over fractions of training subjects. Each vertical dotted line shows a value on the x-axis where the data are plotted. (C) Lineplot showing test performance (accuracy) over scanning sessions. Each vertical dotted line shows a value on the x-axis where the data are plotted. The surfaces represent the 95 % confidence interval.*

## 3.2 Scaling Reserves for Cognition, Age, and Sex Depend on Sample Size and Scanning Time

Figure 4 shows the results of the scaling experiments for cognition, age, and sex using KRR and SVM in a different way to further explore the tradeoff between sample size and scan time.

Figure S1 depicts the same for the elastic-net models. From the heatmaps (panels A, C, E) we can see that for all three of these targets, both the number of training subjects and the number of scanning sessions lead to a higher prediction performance. From this follows that both these variables can be limiting factors, and that both need to be considered when designing the data acquisition scheme in a study. Our results also show that the best prediction accuracies could only be reached at the maximum numbers for both variables, indicating that for these tasks, both a larger sample size and longer scan times might lead to higher prediction values. Still, it matters which parameter is tweaked, as is demonstrated by the line plots showing the $R^2$ over the total amount of scan time the model is trained on (panels B, D, F). For a given number of sessions, adding more subjects could only improve performance by so much, at least for cognition and sex. For those two targets, adding more subjects did not improve prediction accuracy for the lower scan times after a certain point. Age showed more potential improvements, as the curves show the same upwards trend as was already evident in the scaling curves in figure 2.

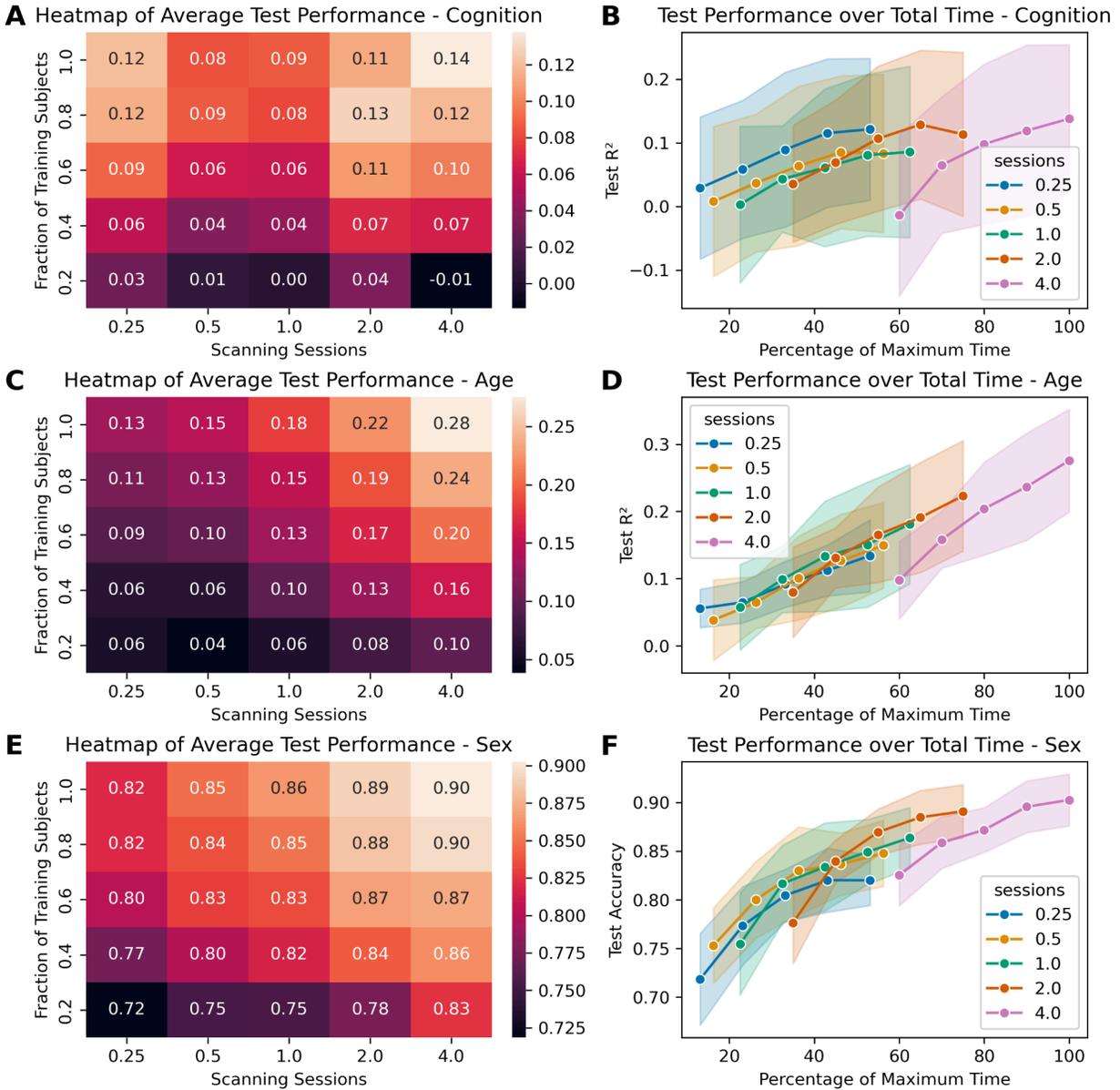

Figure 4: Scaling effects predicting different targets from FC using KRR (for cognition and age) or SVM (for sex) as the model. (A) Heatmap showing the average coefficient of determination ($R^2$) for each combination of fraction of training subjects (rows) and number of scanning sessions (columns) for cognition. (B) Test performance ($R^2$) scaling curve over the maximum available scan time for predicting cognition. 100 % represents four scanning sessions with all available training subjects. Shown is the mean performance for each number of sessions, with the SD as the error band. (C) Heatmap for age. (D) Scaling curve for age. (E) Heatmap for sex (accuracy). (F) Scaling curve for sex (accuracy).

# 4 Discussion

This study investigated the prediction performance and scaling properties of MRI features derived using various methods. Generally, FC-based features did best, with measures of variability and GSP-based metrics coming second, depending on the target. Mean came out the lowest in our comparison, not reaching significance for any target. Similarly, we were not able to predict some targets at all, getting only significant predictions for cognition, age, and sex, but not for mental health, processing speed, and substance use. Generally, KRR and SVM performed similarly or better than the elastic-net models. In terms of scaling, only the better-performing features scaled well, with the other features saturating on a lower performance metric. Only for the former did we observe an upward trend for the highest combinations of sample size and scan time. As implied by the upward trend in the scaling curves, age seemed to have higher reserves than sex and cognition, indicating that this target would benefit most from adding more data. Generally, both increasing scan time and sample size might lead to higher prediction performances. However, these two variables can not be tweaked independently. For a given scan time, adding more subjects could only increase prediction accuracy by so much, before the curve became saturated.

For all targets where prediction was possible, FC and its variants came out as the best-performing features. These results indicate that there seems to be some information about the targets contained in the interactions *between* regional brain signals, which can not be approximated by looking at regional signals in isolation, thus leading to a lower performance in features that quantify variability *within* a region. This finding can be related to the topic of functional integration and functional segregation, a fundamental principle of how the brain functions. Segregation refers to processing within regions, while integration refers to the interaction of brain regions. Viewed in this light, regional features are measures of segregation, while FC is classically viewed as a measure of integration (Friston, 1994). Earlier results found that general cognitive abilities are supported by good integration (Wang et al., 2021), which might also explain why cognition is also better predicted by measures of integration. This framework might also explain why PSD was one of the better-performing features for predicting age, since it is not in the regional space, but instead quantifies the relative contributions of complex patterns of brain activation to the signal, making it another measure of integration. There is still potential for improvement on FC-based measures, as is shown by coupled FC outperforming regular FC when predicting sex, as well as when predicting cognition using elastic-net regression. This could mean that in this case, the graph filtering that was performed as part of calculating decoupled FC had a denoising effect, filtering out information not needed for prediction. While this result needs to be confirmed on other data, it illustrates the potential of graph filtering methods for behavior prediction, which was already demonstrated in one study (Brahim & Farrugia, 2020) and could be explored further in future studies.

Comparing the models, KRR performed either similarly to elastic-net regression (for cognition), or generally better (for age). Equivalently for sex, the SVM generally outperformed the elastic-net classifier, even though in this case this also came with lower performance reserves for the SVM. Both KRR and SVM use the kernel trick, meaning that in contrast to both elastic-net models, they do not calculate a linear combination of the features for their prediction, but instead use the similarity between subjects in terms of the features to calculate a linear combination of the training subjects. This property makes KRR and SVM especially suited for problems when there are few data points (in this case subjects), but high feature dimensionality. This could be the reason for why the kernel-based methods KRR and SVM have similar or better prediction performance in our study. However, this comes at the loss of being able to directly observe feature importance through beta weights, as is possible for the elastic-net models.

Given the success of FC to predict cognition in previous studies, our hope was that other features would also be able to predict different targets. For the features tested in this study, this turned out not to be the case. In the study by Ooi et al. (2022), FC does predict their dissatisfaction and emotion factors to some degree, while here, FC does not predict mental health. However, since their study uses Pearson's correlation as the performance metric, performance might be overestimated, as correlation coefficients are insensitive to errors in bias and scale of the prediction. We instead opted to use $R^2$, as it is a more rigorous measure, potentially causing any small effects to disappear for that reason. Still, the question remains why cognition, age, and sex could be predicted, but mental health, processing speed, and substance use not. One reason for this could be the sampling strategy of the HCP dataset. While in the recruitment it was avoided to have a "supernormal" sample, subjects with neurodevelopmental, psychiatric, and neurologic disorders were still excluded (Van Essen et al., 2012). Given that psychiatric disorders have a lifetime prevalence of around 50 % (Kessler et al., 2005), this excludes a large chunk of the population, and with that much variability in mental health and substance use. Additionally, given that the age range in the dataset is relatively small (22-37), and that processing speed varies with age (Salthouse, 1996, 2000), the variation here is also likely small relative to the range across the whole population. Future studies on large-scale datasets that cover more variability in mental health, substance use, and processing speed might thus shed more light on whether the prediction performance was low because variability in the target measures was lacking, or whether all tested measures of brain activity were unsuited to predict these targets.

Regarding the scaling effects of increasing the number of training subjects and the number of scanning sessions, we found that both were important for predicting inter-individual differences. That both scan time and sample size matter for this aim seems like an obvious conclusion that is still worth repeating, as long as it is claimed that enormous sample sizes are needed for neuroimaging biomarkers to work (Harvey et al., 2023; Marek et al., 2022). Instead, we should focus on the quality of the data as much as we can, in order to make the most efficient use of resources. In this regard, this study supports the assertion that more focus should be put on scan time when designing studies using resting-state fMRI and similar modalities, as has been suggested before (Ooi et al., 2024). Of course, there are other considerations that decide over the quality of a dataset (Makowski et al., 2024), but scan time has been shown to considerably influence the reliability of measures such as FC (Noble et al., 2017). Interestingly, the conclusion that the available scan time limits prediction capabilities did not hold true for age prediction. This might be due to age being related to stable, relatively easily identifiable patterns in brain activity that do not need much scan time for the model to learn. Given the small age range, adding more subjects introduces more variability to the sample, which was more likely the limiting factor in this case. All in all, our results support the conclusion that in order for a model to generalize, i.e. make accurate predictions on unseen data, both sample size and scan time are important factors. Of course, these two parameters drive prediction accuracy in different ways. While adding more subjects adds more variety to the sample, scanning subjects for longer increases the precision with which each feature can be calculated in each subject. As the results displayed in figure 4 show, this can place a limiting factor on how well the model performs.

Our discussion has already alluded to some limitations of this study. One is the limited variability in target measures, as caused by the sampling strategy of the HCP dataset which focused on healthy, young adults. This is problematic when the goal is to predict mental health (including substance use disorders), or variables that change with age, such as processing speed. Future studies should thus repeat these analyses on datasets with more diversity, such as the HCP Aging data (Bookheimer et al., 2019), which has a broader age range, and the healthy brain network data (Alexander et al., 2017), which follows a transdiagnostic approach.

This study compared a variety of measures in their ability to predict various behavioral variables, and assessed how prediction performance scaled with sample size and scan time. It was found that measures of integration, such as FC and its variants, generally outperformed regional measures, which likely did not capture the nuanced interactions of brain regions effectively. While FC remains a strong baseline, the results also point to the potential of more complex measures based on GSP, which should be explored further in the future. The results also reaffirm the conclusion that more emphasis should be put on scanning time when designing data acquisition schemes with the goal to predict inter-individual differences, as it places a limit on prediction performance for some targets. The limited variability in the HCP dataset likely constrained predictions of mental health, processing speed and substance use, pointing to the need for datasets with broader demographic and clinical diversity. In sum, while this study reaffirms the robustness of FC as a predictive measure, it also points toward exciting opportunities for methodological advances and the value of richer, more diverse datasets for behavioral prediction in neuroscience.

## Data and Code Availability

Data are available through the Human Connectome Project, WU-Minn Consortium: https://www.humanconnectome.org/study/hcp-young-adult.
All analysis code can be found in our Github repository:
https://github.com/mschoettner/Comparing-and-Scaling-fMRI-Features-for-Behavior-Prediction

## Author Contributions

M.S. analyzed and visualized the data and wrote the original manuscript. J.P., T.B. and M.S. curated the dataset. T.B., J.P, and P.H. reviewed and edited the manuscript. The work was supervised by P.H. All authors took part in conceptualization.

## Competing Interests

The authors declare no conflicts of interest.

## Acknowledgements

Data were provided by the Human Connectome Project, WU-Minn Consortium (Principal Investigators: David Van Essen and Kamil Ugurbil; 1U54MH091657) funded by the 16 NIH Institutes and Centers that support the NIH Blueprint for Neuroscience Research; and by the McDonnell Center for Systems Neuroscience at Washington University. This work has been supported by Swiss National Science Foundation grant #197787.